\documentclass[traditabstract]{aa}

\usepackage{txfonts}
\usepackage{graphicx}
\usepackage{natbib}
\usepackage{xspace}
\usepackage{url}
\bibpunct{(}{)}{;}{a}{}{,}

\begin{document}

\title{$H_0$ from ten well-measured time delay lenses} 
\titlerunning{$H_0$ from ten well-measured time delay lenses}

\author{S.~Rathna Kumar\inst{\ref{iia},\ref{prl}}
\and C.~S.~Stalin\inst{\ref{iia}}
\and T.~P.~Prabhu\inst{\ref{iia}}
}
\authorrunning{Rathna Kumar et al.}

\institute{
Indian Institute of Astrophysics, II Block, Koramangala, Bangalore 560 034, India \\ \email{rathna@iiap.res.in} \label{iia}
\and
Physical Research Laboratory, Navrangpura, Ahmedabad 380 009, India \label{prl}
}

\date{Received 10 April 2014 / Accepted 27 May 2015}
\abstract{
In this work, we present a homogeneous curve-shifting analysis using the 
difference-smoothing technique of the publicly 
available light curves of 24 gravitationally lensed quasars, for which time 
delays have been reported in the literature. The uncertainty of each measured time delay 
was estimated using realistic simulated light curves. The recipe for generating
such simulated light curves with known time delays in a plausible range around the measured
time delay is introduced here. We identified 14 gravitationally lensed quasars 
that have light curves of sufficiently good quality to enable the measurement 
of at least one time delay between the images, adjacent to each other in terms 
of arrival-time order, to a precision of better than 20\% (including 
systematic errors). We modeled the mass distribution of ten of those systems that have known lens redshifts, accurate astrometric data, and sufficiently 
simple mass distribution, using the publicly available PixeLens code to infer 
a value of $H_0$ of 68.1 $\pm$ 5.9 km s$^{-1}$ Mpc$^{-1}$ (1$\sigma$ 
uncertainty, 8.7\% precision) for a spatially flat universe 
having $\Omega_m$ = 0.3 and $\Omega_\Lambda$ = 0.7. We note here that the lens modeling approach followed in this work is a relatively simple one and does not account for subtle systematics such as those resulting from line-of-sight effects and hence our $H_0$ estimate should be considered as indicative. 
}


\maketitle

\section{Introduction}
The Hubble constant at the present epoch ($H_0$), the current expansion rate 
of the universe, is an important cosmological parameter. All extragalactic 
distances, as well as the age and size of the universe depend on $H_0$. It is also an 
important parameter in constraining the dark energy equation of state and it is used
as input in many cosmological simulations (\citealt{Freedman2010}; 
\citealt{Planck2014}). Therefore, precise estimation of 
$H_0$ is of utmost importance in cosmology.

Estimates of $H_0$ available in the literature cover a wide range of 
uncertainties from $\sim$2\% to $\sim$10\% and the value ranges between 
60 and 75 km s$^{-1}$ Mpc$^{-1}$. The most reliable
measurements of $H_0$ known to date include
\begin{description}
\item[--] the Hubble Space Telescope (HST) Key Project (72 $\pm$ 8 km s$^{-1}$ Mpc$^{-1}$; \citealt{Freedman2001}),
\\
\item[--] the HST Program for the Luminosity Calibration of Type Ia Supernovae by Means of Cepheids 
(62.3 $\pm$ 5.2 km s$^{-1}$ Mpc$^{-1}$; \citealt{Sandage2006}), 
\\
\item[--] Wilkinson Microwave Anisotropy Probe (WMAP) (70.0 $\pm$ 2.2 km s$^{-1}$ Mpc$^{-1}$; \citealt{Hinshaw2013}), 
\\
\item[--] Supernovae and $H_0$ for the Equation of State (SH0ES) Program (73.8 $\pm$ 2.4 km s$^{-1}$ Mpc$^{-1}$; \citealt{Riess2011}),
\\ 
\item[--] Carnegie Hubble Program (CHP) (74.3 $\pm$ 2.6 km s$^{-1}$ Mpc$^{-1}$; \citealt{Freedman2012}),
\\
\item[--] the Megamaser Cosmology Project (MCP) (68.9 $\pm$ 7.1  km s$^{-1}$ Mpc$^{-1}$; \citealt{Reid2013}; \citealt{Braatz2013}),
\\ 
\item[--] Planck measurements of the cosmic microwave background (CMB) anisotropies (67.3 $\pm$ 1.2  km s$^{-1}$ Mpc$^{-1}$; \citealt{Planck2014}), and
\\
\item[--] Strong lensing time delays (75.2$^{+4.4}_{-4.2}$ km s$^{-1}$ Mpc$^{-1}$; \citealt{Suyu2013}).
\end{description}
It is worth noting here that the small uncertainties in $H_0$ measurements resulting from WMAP and Planck  crucially depend on the assumption of a spatially flat universe.

Although the values of $H_0$ obtained from different methods are 
consistent with each other within 2$\sigma$ given the current level of precision, 
all of the above methods of determination of $H_0$ suffer from systematic
uncertainties. Therefore as the measurements increase in precision, multiple approaches based on different physical principles 
need to be pursued so as to be able to identify unknown 
systematic errors present in any given approach.  

The phenomenon of strong gravitational lensing offers an elegant method to 
measure $H_0$. 
For gravitationally lensed sources that show variations
in flux with time, such as  quasars, it is possible to measure the time delay between the various
images of the background source.  The time delay,
which is a result of the travel times for photons being different along the 
light paths corresponding to the lensed images, has two origins: (i) the geometric difference 
between the light paths and (ii) gravitational delay due
to the dilation of time as photons pass in the vicinity of the lensing mass.
Time delays, therefore depend on the cosmology, through the distances between
the objects involved, and on the radial mass profile of the lensing galaxies.
This was shown theoretically five decades ago by \citet{Refsdal1964} long 
before the discovery of the first gravitational lens Q0957+561 by \citet{Walsh1979}.

Estimation of $H_0$ through gravitational lens time delays, although it has its
own degeneracies, is based on the well-understood physics of General Relativity, 
and compared to distance
ladder methods, is free from various calibration issues. In addition to 
measuring $H_0$, measurement of time delays between the light curves of 
a lensed quasar can be used to study the microlensing variations present 
in the light curves, and to study the structure of the quasar 
(\citealt{Hainline2013}; \citealt{Mosquera2013}). However, these time delay measurements of $H_0$ are extremely challenging
because of the need of an intensive monitoring program that offers 
high cadence and good-quality photometric data over a long period of time. This type of program would then be able to cope with the presence of 
uncorrelated variations present in the lensed quasar lightcurves, which can interestingly arise due to microlensing by stars in the lensing galaxy \citep{Chang1979} or for mundane reasons, such as the presence of additive flux shifts in the photometry \citep{Tewes2013a}. 
Moreover, the estimation of $H_0$ from such high-quality data is hampered
by the uncertainty on lens models. Recently, using time delay measurements from high-quality optical and radio light curves, deep and high-resolution 
imaging observations of the lensing galaxies and lensed AGN host galaxy, and the measurement of stellar velocity dispersion of the lens galaxy to perform detailed modeling, 
\citet{Suyu2013} report a $H_0$ of 
75.2$^{+4.4}_{-4.2}$ km s$^{-1}$ Mpc$^{-1}$ through the study of two 
gravitational lenses namely RX J1131$-$1231 and CLASS B1608+656.

Another approach is to perform simple modeling of a relatively large sample of gravitational lenses with moderate-precision time 
delay measurements. In this way, it should be possible
to obtain a precise determination of the global value of
$H_0$, even if the $H_0$ measurements from individual lenses have large uncertainties. In addition,  when inferring $H_0$ from a relatively large sample of lenses, line-of-sight effects that bias the $H_0$ measurements from individual lenses \citep[see][Sect. 2]{Suyu2013} should tend to average out, although a residual systematic error must still remain \citep{Hilbert2007, Fassnacht2011}.
A pixelized method of lens modeling is available in the literature and
is also implemented in the publicly available code PixeLens \citep{Saha2004}.
Using this code, \citet{Saha2006} have found $H_0$ = 
72$^{+8}_{-11}$ km s$^{-1}$ Mpc$^{-1}$ for a sample of ten time delay lenses. Performing 
a similar analysis on an extended sample of 18 lenses \citet{Paraficz2010} 
obtained $H_0$ = 66$^{+6}_{-4}$ km s$^{-1}$ Mpc$^{-1}$. Here, we present an 
estimate of $H_0$ using the pixellated modeling approach
on a sample of carefully selected lensed quasars. 
So far, time delays have been reported for 
24 gravitationally lensed quasars among the hundreds of such strongly lensed 
quasars known. However, the quality of the light curves and the techniques used 
to infer these time delays vary between systems. In this work, we 
apply the difference-smoothing technique, introduced in \citet{Kumar2013}, to 
the publicly available light curves of the 24 systems in a homogeneous manner,
first to cross-check the previously measured time delays and then to select a subsample of suitable lens
systems to determine $H_0$. 

The paper is organized as follows. Section \ref{section:curve-shifting} 
describes the technique used for time delay determination and introduces a 
recipe for creating realistic simulated light curves with known time delays; 
the simulated light curves are used in this work to estimate the uncertainty of each measured delay.
In Sect. \ref{section:application}, the application of the curve-shifting 
procedure to the 24 systems is described. In Sect. 
\ref{section:lens-modelling}, we infer $H_0$ from the lens-modeling of those 
systems that have at least one reliably measured time delay, known lens redshift, 
accurate astrometric data, and sufficiently simple mass distribution. We 
conclude in Sect. \ref{section:conclusion}. 

\section{Time delay determination}
\label{section:curve-shifting}
In this section, we briefly describe the previously reported difference-smoothing 
technique,  which contains
one modification to the original version  (see \citealt{Kumar2013} for  details). We 
then introduce a recipe for simulating realistic light curves having known 
time delays in a plausible range around the measured delay in order to 
estimate its uncertainty. We also present an approach for tuning the free 
parameters of the difference-smoothing technique for a given dataset.

\subsection{Difference-smoothing technique}
A$_i$ and B$_i$ are the observed magnitudes constituting light curves A and B sampled at epochs $t_i$ ($i=1,2,3,...,N$). Light curve A is selected as the reference. We shift  light curve B in time with respect to  light curve A by an amount $\tau$.  This shifted version B$^\prime$ of B is given by
\begin{eqnarray}
\mathrm{B}_i^\prime&=&\mathrm{B}_i, \\
t_i^\prime&=&t_i + \tau.
\end{eqnarray}
We note here that we do not apply any flux shift to light curve B as in \citet{Kumar2013}, since we have found that doing so considerably increases the computational time without significantly changing the results. 

For any given estimate of the time delay $\tau$, we form a difference light curve having points $d_i$ at epochs $t_i$,
\begin{equation}
\label{equation:diffcurve}
d_i(\tau) = \mathrm{A}_i- \frac{\sum_{j=1}^N w_{ij} \mathrm{B}_j^\prime}{\sum_{j=1}^N w_{ij}}, 
\end{equation}
where the weights $w_{ij}$ are given by
\begin{equation}
\label{equation:pairing}
w_{ij} = \frac{1}{\sigma_{\mathrm{B}_j}^2} e^{-(t_j^\prime-t_i)^2 / 2\delta^2}. 
\end{equation}
The parameter $\delta$ is the decorrelation length and $\sigma_{\mathrm{B}_j}$ denotes the photometric error of the magnitude B$_j$. We calculate the uncertainty of each $d_i$ as
\begin{equation}
\sigma_{d_i} = \sqrt{\sigma_{\mathrm{A}_i}^2+\frac{1}{\sum_{j=1}^N w_{ij}}},
\end{equation}
where $w_{ij}$ are given by Eq. \ref{equation:pairing}. 

We now smooth the difference curve $d_i$ using a Gaussian kernel to obtain a model $f_i$ for the differential extrinsic variability
\begin{equation}
f_i = \frac{\sum_{j=1}^N \nu_{ij} \, d_j}{\sum_{j=1}^N \nu_{ij}},
\end{equation} 
where the weights $\nu_{ij}$ are given by
\begin{equation}
\label{equation:smoothing}
\nu_{ij} = \frac{1}{\sigma_{d_j}^2}{e^{-(t_j-t_i)^2 / 2s^2}}.
\end{equation}
The smoothing time scale $s$ is another free parameter of this method. The uncertainty of each $f_i$ is computed as
\begin{equation}
\sigma_{f_i} = \sqrt{\frac{1}{\sum_{j=1}^N \nu_{ij}}}.
\end{equation} 

We optimize  the time delay estimate $\tau$ to minimize the residuals between the difference curve $d_i$ and the much smoother $f_i$. To quantify the mismatch between $d_i$ and $f_i$, we define a normalized $\chi^2$, 
\begin{equation}
\label{chi2}
\overline{\chi}^2 = \left[ \sum_{i=1}^N \frac{(d_i-f_i)^2}{\sigma_{d_i}^2+\sigma_{f_i}^2} \right] / \left[ \sum_{i=1}^N \frac{1}{\sigma_{d_i}^2+\sigma_{f_i}^2} \right],  
\end{equation}
and minimize this $\overline{\chi}^2(\tau)$ using a global optimization. 

In the above description, since light curves A and B are not interchangeable, we systematically perform all computations for both permutations of A and B, and minimize the average of the two resulting values of $\overline{\chi}^2$.

\subsection{Simulation of light curves}
In \citet{Kumar2013}, in order to estimate the uncertainty of the time delay measured using the difference-smoothing technique, we made use of realistic simulated light curves, which were created following the procedure introduced in \citet{Tewes2013a}. In this work, we introduce an independent recipe for creating simulated light curves.

We infer the underlying variation $\mathrm{A}(t)$ of the light curve A at the epoch $t_i$ based on the magnitudes A$_j$ for all the epochs as
\begin{equation}
\mathrm{A}(t_i) = \frac{\sum_{j=1}^N \frac{1}{\sigma_{\mathrm{A}_j}^2} e^{-(t_j-t_i)^2 / 2m^2} \mathrm{A}_j}{\sum_{j=1}^N \frac{1}{\sigma_{\mathrm{A}_j}^2} e^{-(t_j-t_i)^2 / 2m^2}},  
\end{equation}
where the value of $m$ is set to equal the mean sampling of the light curves calculated after excluding the large gaps following
a 3$\sigma$ rejection criterion. For those points having the nearest neighboring points on both sides separated by a value less than or equal to $m$, we compute the values of $(\mathrm{A_i}-\mathrm{A}(t_i))/\sigma_{\mathrm{A}_i}$, the standard deviation of which is multiplied to the error bars $\sigma_{\mathrm{A}_i}$ to obtain the rescaled error bars $\hat{\sigma}_{\mathrm{A}_i}$. We note here that the rescaling is applied for all the epochs and not just the epochs of points used in computing the rescaling factor. Similarly for the B light curves the rescaled error bars $\hat{\sigma}_{\mathrm{B}_i}$ are obtained. This rescaling inferred from the local scatter properties of the light curves is done because the magnitudes of the original error bars may suffer from systematic underestimation or overestimation. 

We merge  light curves A and B by shifting the B light curve by the time delay found ($\Delta t$) and subtracting the differential extrinsic variability $f_i$ corresponding to the delay from the A light curve. This merged light curve $\mathrm{M}_i$, whose errors we denote $\sigma_{\mathrm{M}_i}$, consists of the magnitudes $\mathrm{A}_i-f_i$  at times $t_i$ and having errors $\hat{\sigma}_{\mathrm{A}_i}$ and the magnitudes $\mathrm{B}_i$ at times $t_i+\Delta t$ and having errors $\hat{\sigma}_{\mathrm{B}_i}$. We now model the quasar brightness variation $\mathrm{M}(t)$ as
\begin{equation}
\mathrm{M}(t) = \frac{\sum_{j=1}^{2N} \frac{1}{\sigma_{\mathrm{M}_j}^2} e^{-(t_j-t)^2 / 2m^2} \mathrm{M}_j}{\sum_{j=1}^{2N} \frac{1}{\sigma_{\mathrm{M}_j}^2} e^{-(t_j-t)^2 / 2m^2}}. 
\end{equation}

We then model the quasar brightness variation using only the $\mathrm{A}$ points in $\mathrm{M}_i$ as
\begin{equation}
\mathrm{M_A}(t) = \frac{\sum_{j=1}^{N} \frac{1}{\hat{\sigma}_{\mathrm{A}_j}^2} e^{-(t_j-t)^2 / 2m^2} (\mathrm{A}_j-f_j)}{\sum_{j=1}^{N} \frac{1}{\hat{\sigma}_{\mathrm{A}_j}^2} e^{-(t_j-t)^2 / 2m^2}} 
\end{equation}
and only the $\mathrm{B}$ points in $\mathrm{M}_i$ as
\begin{equation}
\mathrm{M_B}(t) = \frac{\sum_{j=1}^{N} \frac{1}{\hat{\sigma}_{\mathrm{B}_j}^2} e^{-(t_j+\Delta t-t)^2 / 2m^2} \mathrm{B}_j}{\sum_{j=1}^{N} \frac{1}{\hat{\sigma}_{\mathrm{B}_j}^2} e^{-(t_j+\Delta t-t)^2 / 2m^2}}.
\end{equation}
The residual extrinsic variations present in the A and B light curves can now be calculated as
\begin{equation}
f_{\mathrm{A}_i} = \mathrm{M_A}(t_i)-\mathrm{M}(t_i)
\end{equation}
and
\begin{equation}
f_{\mathrm{B}_i} = \mathrm{M_B}(t_i)-\mathrm{M}(t_i).
\end{equation}

We can now simulate light curves $\mathrm{A}_i^{simu}$ and $\mathrm{B}_i^{simu}$ having a time delay of $\Delta t+dt$ between them by sampling $\mathrm{M}(t)$ at appropriate epochs and adding terms for extrinsic variations and noise,
\begin{equation}
\mathrm{A}_i^{simu} = \mathrm{M}\left(t_i-\frac{dt}{2}\right) + f_i + f_{\mathrm{A}_i} + N^*(0,1)\hat{\sigma}_{\mathrm{A}_i}
\end{equation}
and
\begin{equation}
\mathrm{B}_i^{simu} = \mathrm{M}\left(t_i+\Delta t+\frac{dt}{2}\right) + f_{\mathrm{B}_i} + N^*(0,1)\hat{\sigma}_{\mathrm{B}_i},
\end{equation}
where $N^*(0,1)$ is a random variate drawn from a normal distribution having mean 0 and variance 1. These simulated light curves are then assigned the times $t_i$ and the error bars $\sigma_{\mathrm{A}_i}$ and $\sigma_{\mathrm{B}_i}$ for the A and B light curves, respectively. Including the terms $f_{\mathrm{A}_i}$ and $f_{\mathrm{B}_i}$ in the calculation of $\mathrm{A}_i^{simu}$ and $\mathrm{B}_i^{simu}$, respectively, ensures that our simulated light curves contain extrinsic variability on all time scales, just as in the real light curves. 

Here again in the above description, since light curves A and B are not interchangeable, we systematically perform all computations for both permutations of A and B, and average the corresponding values of $\mathrm{A}_i^{simu}$ and $\mathrm{B}_i^{simu}$, before adding the noise terms. 

\subsection{Choice of free parameters}
The value chosen for  the decorrelation length $\delta$  needs to be  equivalent to the temporal sampling of the light curves. In 
this work, we set $\delta$ equal to $m$, the mean sampling of the light curves calculated after excluding
the large gaps following a 3$\sigma$ rejection criterion.

The value chosen for the smoothing time scale $s$ needs to be  significantly larger 
than $\delta$. In this work, its value is optimized such that the larger of the maximum absolute values of 
$\frac{f_{\mathrm{A}_i}}{\hat{\sigma}_{\mathrm{A}_i}}$ and 
$\frac{f_{\mathrm{B}_i}}{\hat{\sigma}_{\mathrm{B}_i}}$, which quantify the 
residual extrinsic variations in units of photometric noise 
for the A and B light curves respectively, is equal to 2. This choice 
ensures that the value of $s$ is small enough to adequately model the 
extrinsic variations, so that the extreme values of residual extrinsic 
variations are not significantly larger than the noise in the data.  

Again as in the above description, because light curves A and B are not 
interchangeable, we systematically perform all the computations for both 
permutations of A and B, and average the corresponding maximum absolute values. 

\subsection{Estimation of uncertainty}
We create 200 simulated light curves having a true delay of $\Delta t$ between 
them. The difference-smoothing technique is applied on each of them to obtain 
200 delay values. The standard deviation of the 200 delay values gives us the 
random error, and the systematic error is obtained by the difference between 
the mean of the 200 delay values and the true delay. The total error 
$\Delta \tau_0$ is obtained by adding the random error and the systematic 
error in quadrature. 

However, as noted by \citet{Tewes2013a}, it is important to simulate light 
curves that have not only the time delay $\Delta t$ found, but also other time 
delays in a plausible range around $\Delta t$, so as to obtain a reliable 
estimate of the uncertainty (see also Sect. 3.2 in \citealt{Kumar2013}). To 
this end, we also simulate 200 light curves for each true delay that differs 
from $\Delta t$ by 
$\pm\Delta \tau_0, \pm(\Delta \tau_0 + \Delta \tau_1), ... \, , \pm(\Delta \tau_0 + \Delta \tau_1 + ... + \Delta \tau_{n-1})$, in each step updating the total error $\Delta \tau_n$ by adding the maximum obtained 
value of the random error and the maximum obtained absolute value of the 
systematic error in quadrature. The value of $n$ is chosen to be the smallest integer for 
which 
\begin{equation}
\Delta \tau_0 + \Delta \tau_1 + ... + \Delta \tau_{n-1} \ge 2\Delta \tau_n.
\end{equation} 
This ensures that we have 
simulated light curves over a range of delay values that is at least as wide as 
or wider than the 95.4\% confidence interval implied by the stated final error 
$\Delta \tau_n$.      

\subsection{Testing the robustness of the procedure}
\begin{figure*}[htbp]
\begin{center}
\resizebox{1.0 \hsize}{!}{\includegraphics{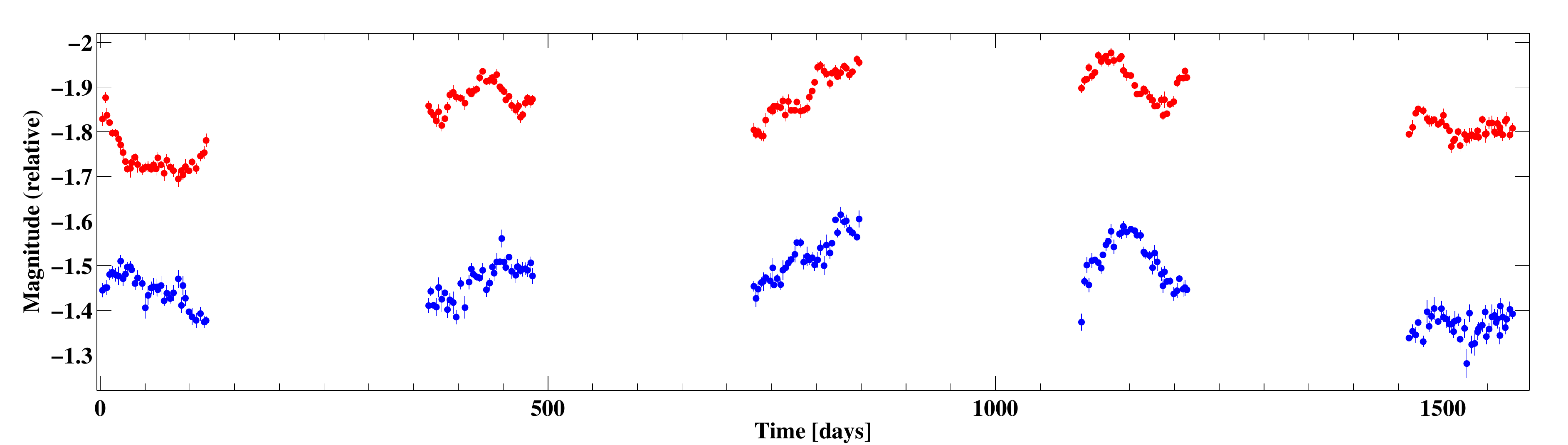}}
\caption{Light curves from the Strong Lens Time Delay Challenge file ``tdc1\_rung3\_quad\_pair9A.txt''. Light curve A is shown in red and light curve B in blue.}
\label{figure:lightcurves}
\end{center}
\end{figure*}

\begin{figure*}[htbp]
\begin{center}
\resizebox{1.0 \hsize}{!}{\includegraphics{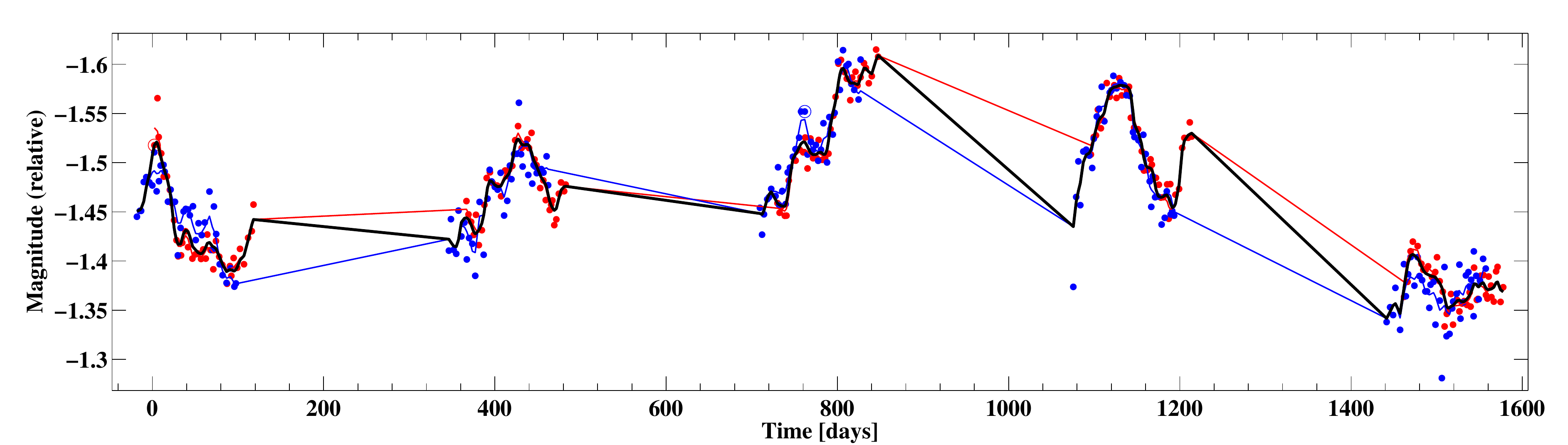}}
\caption{Light curves A and B  from Fig. \ref{figure:lightcurves} have been merged, with  light curve A as reference, after shifting  light curve B by the measured time delay of $\Delta t$ = $-$20.5 days and subtracting the differential extrinsic variability from  A. $M_A(t)$ sampled at the epochs $t_i$ and $M_B(t)$ sampled at the epochs $t_i+\Delta t$ are connected by red and blue lines, respectively. $M(t)$ sampled at the epochs $t_i$ and $t_i+\Delta t$ are connected by black lines. The optimum free parameters for this pair of light curves were found to be $\delta$ = 3.1 days and $s$ = 139.0 days. The magnitudes at those epochs corresponding to maximum absolute values of $\frac{f_{\mathrm{A}_i}}{\hat{\sigma}_{\mathrm{A}_i}}$ and $\frac{f_{\mathrm{B}_i}}{\hat{\sigma}_{\mathrm{B}_i}}$ have been circled. The negative value of time delay implies that  light curve A leads  light curve B. The magnitudes are shown without error bars for convenience of display.}
\label{figure:merged lightcurves}
\end{center}
\end{figure*}

\begin{figure}
\begin{center}
\resizebox{1.0 \hsize}{!}{\includegraphics{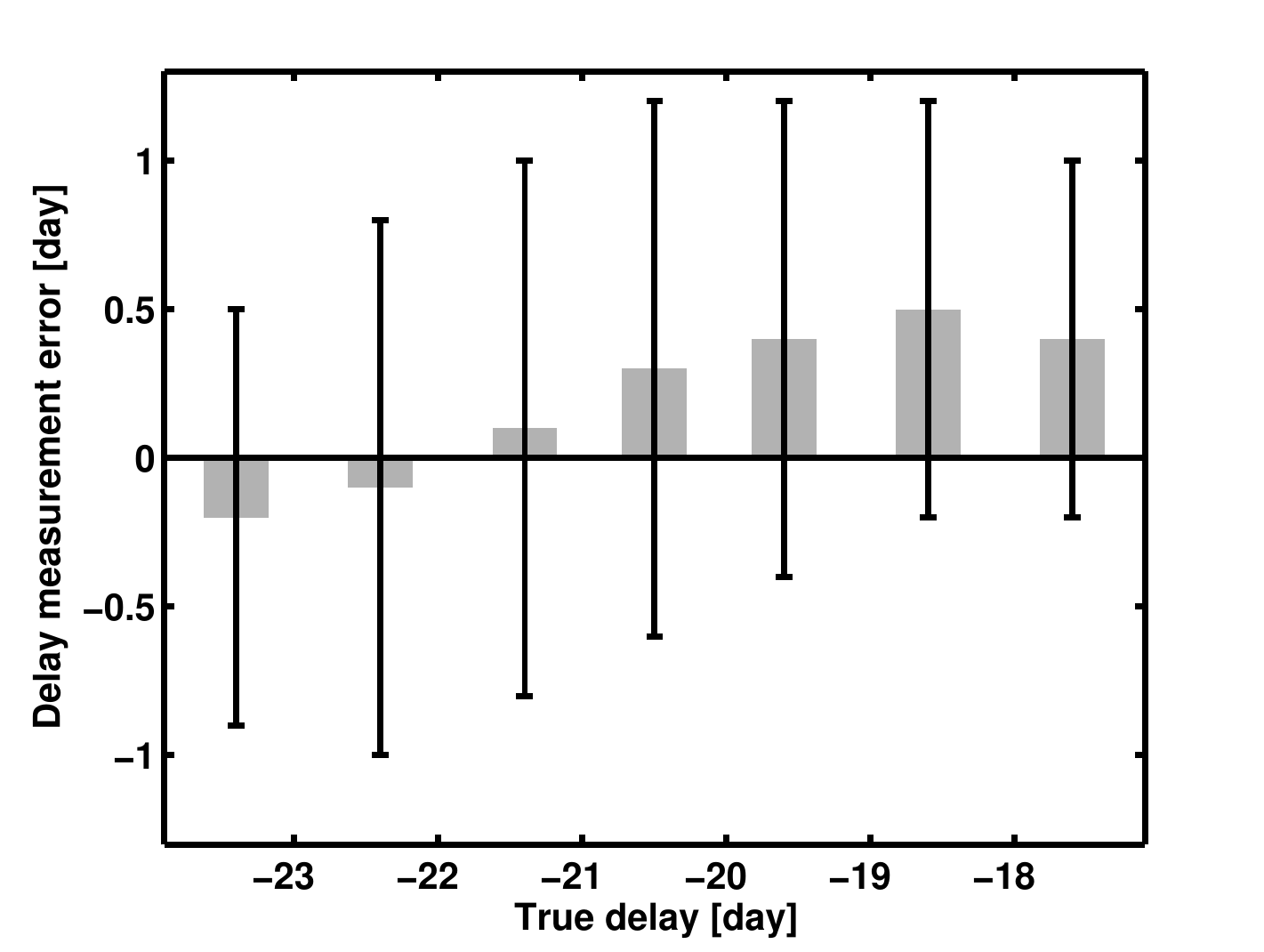}}
\caption{Error analysis of the time delay measurement based on delay estimations on simulated light curves that mimic the light curves displayed in Fig. \ref{figure:lightcurves}. The horizontal axis corresponds to the value of the true time delay used in these simulated light curves. The gray colored rods and 1$\sigma$ error bars show the systematic biases and random errors, respectively. Our measured time delay of $\Delta t$ = $-$20.5 $\pm$ 1.0 days is discrepant with the true time delay of 22.75 days listed in the TDC1 truth files at the level of 2.25$\sigma$. The difference in sign of time delay is simply a matter of convention.}
\label{figure:error analysis}
\end{center}
\end{figure}

In order to test the robustness of our procedure for estimating the time delay and its uncertainty, we made use of synthetic light curves from the TDC1 stage of the Strong Lens Time Delay Challenge\footnote{\url{http://timedelaychallenge.org/}} \citep{Liao2015}, which are arranged in five rungs having different sampling properties \citep[see][Table 1]{Liao2015}. We applied our procedure on a sample of 250 light curves, 50 from each rung, selected such that we were able to reliably measure time delays from them. Comparing our results with the truth files, we found that all the measured delays agreed with the true delays to within twice the estimated uncertainties, except in one case. For the exceptional case, the discrepancy between the measured delay and the true delay was found to be 2.25$\sigma$. This is still a reasonably good level of agreement, thus demonstrating the robustness of our procedure. We note here that this property of robustness also depends  on the careful choice of free parameters as presented here. For instance, setting $\delta$ equal to the mean sampling of the light curves computed without excluding the large gaps was found to lead to biased time delay measurements, which was especially noticeable for light curves having shorter seasons and larger cadence. We show some plots for the pair of TDC1 light curves corresponding to the exceptional case mentioned above in Figs. \ref{figure:lightcurves}--\ref{figure:error analysis}.     

\section{Time delays of 24 gravitationally lensed quasars}
\label{section:application}

\begin{table*}
\caption{Summary of time delay measurements.}
\label{table:time-delays}
\begin{center}
\begin{tabular}{l c c c c}
\hline
\hline
Object (Reference for data) & Wavebands & Time delay & Reported value\tablefootmark{a} & Our measurement\tablefootmark{a} \\
                            &          &            & (days)                          & (days) \\
\hline
Q0142$-$100 \citep{Koptelova2012} & $R$ & $\Delta t_{AB}$ & 89 $\pm$ 11 & ? \\
\hline
JVAS B0218+357 \citep{Cohen2000} & 8 GHz, 15 GHz & $\Delta t_{AB}$ & 10.1$^{+1.5}_{-1.6}$ (95\% CI) & 10.7 $\pm$ 0.8 \\
\hline
HE 0435$-$1223 (\citealp{Courbin2011}, & $R$ & $\Delta t_{AB}$ & 8.4 $\pm$ 2.1    & 9.8 $\pm$ 1.1 \\
               \citealp{Blackburne2014})                    &     & $\Delta t_{AC}$ & 0.6 $\pm$ 2.3    & 3.1 $\pm$ 2.2 \\
                                   &     & $\Delta t_{AD}$ & 14.9 $\pm$ 2.1   & 13.7 $\pm$ 1.0 \\
                                   &     & $\Delta t_{BC}$ & $-$7.8 $\pm$ 0.8 & $-$8.0 $\pm$ 1.0 \\
                                   &     & $\Delta t_{BD}$ & 6.5 $\pm$ 0.7    & 6.2 $\pm$ 1.5 \\
                                   &     & $\Delta t_{CD}$ & 14.3 $\pm$ 0.8   & 13.6 $\pm$ 0.8 \\
\hline
SBS 0909+532 (\citealp{Goicoechea2008}, & $r$ & $\Delta t_{AB}$ & $-$50$^{+2}_{-4}$ & $-$45.9 $\pm$ 3.1 \\
             \citealp{Hainline2013})    &     &                 &                   &   \\
\hline
RX J0911.4+0551 \citep{Hjorth2002} & $I$ & $\Delta t_{(A1+A2+A3)B}$ & $-$146 $\pm$ 8 (2$\sigma$) & $-$141.9 $\pm$ 12.3 \\
\hline
FBQ 0951+2635 \citep{Jakobsson2005} & $R$ & $\Delta t_{AB}$ & 16 $\pm$ 2 & 7.8 $\pm$ 14.0 \\
\hline
Q0957+561 \citep{Shalyapin2012} & $r$, $g$ & $\Delta t_{AB}$ & 417.4 $\pm$ 0.9 & 420.0 $\pm$ 1.4 \\
\hline
SDSS J1001+5027 \citep{Kumar2013} & $R$ & $\Delta t_{AB}$ & 119.3 $\pm$ 3.3 & 119.7 $\pm$ 1.8 \\
\hline
SDSS J1004+4112 (\citealp{Fohlmeister2007}, & $R$, $r$ & $\Delta t_{AB}$ & $-$40.6 $\pm$ 1.8 & $-$37.2 $\pm$ 3.1 \\
                \citealp{Fohlmeister2008})  &          & $\Delta t_{AC}$ & $-$821.6 $\pm$ 2.1 & $-$822.5 $\pm$ 7.4 \\
                                            &          & $\Delta t_{BC}$ &                    & $-$777.1 $\pm$ 9.2 \\
\hline
SDSS J1029+2623 \citep{Fohlmeister2013} & $r$ & $\Delta t_{A(B+C)}$ & 744 $\pm$ 10 (90\% CI) & 734.3 $\pm$ 3.8 \\
\hline
HE 1104$-$1805 \citep{Poindexter2007} & $R$, $V$ & $\Delta t_{AB}$ & $-$152.2$^{+2.8}_{-3.0}$ & $-$157.1 $\pm$ 3.6 \\
\hline
PG 1115+080 \citep{Tsvetkova2010} & $R$ & $\Delta t_{(A1+A2)B}$ & 4.4$^{+3.2}_{-2.5}$   & 8.7 $\pm$ 3.6 \\
                                  &     & $\Delta t_{(A1+A2)C}$ & $-$12$^{+2.5}_{-2.0}$ & $-$12.1 $\pm$ 3.6 \\
                                  &     & $\Delta t_{BC}$ & $-$16.4 $^{+3.5}_{-2.5}$    & $-$23.9 $\pm$ 5.7 \\
\hline
RX J1131$-$1231 \citep{Tewes2013b} & $R$ & $\Delta t_{AB}$ & 0.7 $\pm$ 1.0    & 0.0 $\pm$ 0.6 \\
                                   &     & $\Delta t_{AC}$ & 0.0 $\pm$ 1.3    & $-$1.1 $\pm$ 0.8 \\
                                   &     & $\Delta t_{AD}$ & 90.6 $\pm$ 1.4   & 91.7 $\pm$ 0.7 \\
                                   &     & $\Delta t_{BC}$ & $-$0.7 $\pm$ 1.5 & $-$1.4 $\pm$ 1.6 \\
                                   &     & $\Delta t_{BD}$ & 91.4 $\pm$ 1.2   & 92.4 $\pm$ 1.4 \\
                                   &     & $\Delta t_{CD}$ & 91.7 $\pm$ 1.5   & 91.3 $\pm$ 1.3 \\
\hline
SDSS J1206+4332 \citep{Eulaers2013} & $R$ & $\Delta t_{AB}$ & 111.3 $\pm$ 3 & 110.3 $\pm$ 1.9 \\
\hline
H1413+117 \citep{Goicoechea2010} & $r$ & $\Delta t_{AB}$ & $-$17 $\pm$ 3 & $-$14.3 $\pm$ 5.5 \\
                                 &     & $\Delta t_{AC}$ & $-$20 $\pm$ 4 & $-$19.9 $\pm$ 10.9 \\
                                 &     & $\Delta t_{AD}$ & 23 $\pm$ 4    & 24.0 $\pm$ 6.8 \\
                                 &     & $\Delta t_{BC}$ &               & ? \\
                                 &     & $\Delta t_{BD}$ &               & ? \\
                                 &     & $\Delta t_{CD}$ &               & 28.6 $\pm$ 9.4 \\
\hline
JVAS B1422+231 \citep{Patnaik2001} & 15 GHz & $\Delta t_{AB}$ & $-$1.5 $\pm$ 1.4 & 1.1 $\pm$ 2.1 \\
                                   &        & $\Delta t_{AC}$ & 7.6 $\pm$ 2.5 & $-$0.4 $\pm$ 3.0 \\
                                   &        & $\Delta t_{BC}$ & 8.2 $\pm$ 2.0 & $-$0.4 $\pm$ 3.2 \\
\hline
SBS 1520+530 \citep{Burud2002b} & $R$ & $\Delta t_{AB}$ & 130 $\pm$ 3 & 124.2 $\pm$ 8.1 \\
\hline
CLASS B1600+434 \citep{Burud2000} & $I$ & $\Delta t_{AB}$ & 51 $\pm$ 4 (95\% CI) & ? \\
CLASS B1600+434 \citep{Koopmans2000} & 8.5 GHz & $\Delta t_{AB}$ & 47$^{+5}_{-6}$    & ? \\
\hline
CLASS B1608+656 (\citealp{Fassnacht1999}, & 8.5 GHz & $\Delta t_{AB}$ & $-$31.5$^{+2.0}_{-1.0}$ & $-$32.4 $\pm$ 3.0 \\
                \citealp{Fassnacht2002})  &         & $\Delta t_{AC}$ &                         & 2.3 $\pm$ 1.2 \\
                                          &         & $\Delta t_{AD}$ &                         & 45.7 $\pm$ 0.9 \\
                                          &         & $\Delta t_{BC}$ & 36.0$^{+1.5}_{-1.5}$ & 37.1 $\pm$ 1.9 \\
                                          &         & $\Delta t_{BD}$ & 77.0$^{+2.0}_{-1.0}$ & 77.6 $\pm$ 3.5 \\
                                          &         & $\Delta t_{CD}$ &                      & 41.3 $\pm$ 1.6 \\
\hline
SDSS J1650+4251 \citep{Vuissoz2007} & $R$ & $\Delta t_{AB}$ & 49.5 $\pm$ 1.9 & 59.2 $\pm$ 15.9 \\
\hline
PKS 1830$-$211 \citep{Lovell1998} & 8.6 GHz & $\Delta t_{AB}$ & 26$^{+4}_{-5}$ & 28.6 $\pm$ 8.0 \\
\hline
WFI J2033$-$4723 \citep{Vuissoz2008} & $R$ & $\Delta t_{AB}$ & $-$35.5 $\pm$ 1.4      & $-$37.6 $\pm$ 2.1 \\
                                     &     & $\Delta t_{AC}$ &                        & 23.6 $\pm$ 2.5 \\
                                     &     & $\Delta t_{BC}$ & 62.6$^{+4.1}_{-2.3}$ & 65.4 $\pm$ 4.3 \\
\hline
HE 2149$-$2745 \citep{Burud2002a} & $V$ & $\Delta t_{AB}$ & 103 $\pm$ 12 & 72.6 $\pm$ 17.0 \\
\hline
HS 2209+1914 \citep{Eulaers2013} & $R$ & $\Delta t_{AB}$ & $-$20.0 $\pm$ 5 & $-$22.9 $\pm$ 5.3 \\
\hline
\end{tabular}
\end{center}
\tablefoot{
\tablefoottext{a}{A negative value of time delay implies that the arrival-time order is the reverse of what is implied in the subscript to $\Delta t$.}
}
\end{table*}

Time delays have been reported for 24 gravitationally lensed quasars. However, the quality of the data and the curve-shifting procedure followed differs from system to system. In this section, we present a homogeneous analysis of their publicly available light curves following the procedure described in the previous section, with the aim of identifying those systems that have reliable time delay measurements. In the case of systems with more than two images, we measured the time delays between all pairs of light curves. The results are summarized in Table \ref{table:time-delays}. All quoted uncertainties are 1$\sigma$ error bars, unless stated otherwise. Additional information on some systems listed in Table \ref{table:time-delays} and discussion on the possible reasons for our inability to reliably measure some of the time delays follow. For all the other systems, our time delay measurements agree with the previously reported values to within 2$\sigma$. 

\begin{description}
\item[--] {\it Q0142$-$100 (UM673):} We were unable to make a reliable time delay measurement using the light curves presented in \citet{Koptelova2012}. This is not surprising given that the light curves are characterized by large seasonal gaps and there are no clear variability features that could be matched between the A and B light curves.
\\
\item[--] {\it JVAS B0218+357:} From 8 GHz and 15 GHz VLA observations reported by \citet{Cohen2000}, we measured time delays of 10.4 $\pm$ 1.0 days and 11.4 $\pm$ 1.5 days, respectively. Taking the weighted average of the two results, we find the time delay to be 10.7 $\pm$ 0.8 days. We note here that \citet{Biggs1999} monitored this system using VLA during the same period as \citet{Cohen2000} at the same two frequencies and  report a time delay of 10.5 $\pm$ 0.4 days (95\% CI).
\\
\item[--] {\it HE 0435$-$1223:} We made use of the light curves presented in \citet{Courbin2011} spanning seven seasons using data from Euler, Mercator, Maidanak, and SMARTS and the light curves presented in \citet{Blackburne2014} spanning eight seasons using data from SMARTS. The SMARTS data used by \citet{Courbin2011} is the same as the first two seasons of data presented in \citet{Blackburne2014}. Hence we excluded the SMARTS data points from the light curves of \citet{Courbin2011} to make it independent of the light curves of \citet{Blackburne2014}. Owing to the differences in the approaches followed by these two teams of authors to derive photometry and also the photometric uncertainties, we avoided merging the two datasets. Our time delay measurements listed in Table \ref{table:time-delays} are the weighted averages of the time delays measured from the two independent sets of light curves. The reported time delay values in Table \ref{table:time-delays} are from \citet{Courbin2011}. The best-fit time delay values reported without uncertainties by \citet{Blackburne2014} are consistent with the values of \citet{Courbin2011} to within 1$\sigma$. In Table \ref{table:HE0435}, we present our measurements of the time delays of HE 0435$-$1223 from the two independent sets of light curves and the resulting weighted averages. For each pair of quasar images, we see that the time delays measured from the two datasets agree to within 2$\sigma$. 
\\
\item[--] {\it SBS 0909+532:} For our analysis, we used only the r-band
data points obtained using the Liverpool Robotic Telescope
between 2005 January and 2007 January presented in
\citet{Goicoechea2008} and \citet{Hainline2013}, based
on homogeneity and sampling considerations.
\\
\item[--] {\it RX J0911.4+0551:} We used the light curves presented in \citet{Hjorth2002}, which were made publicly available by \citet{Paraficz2006}.
\\
\item[--] {\it FBQ 0951+2635:} We used the light curves presented in \citet{Jakobsson2005}, which were made publicly available by \citet{Paraficz2006}.
\\
\item[--] {\it Q0957+561:} From the $r$-band and $g$-band light curves presented in \citet{Shalyapin2012}, we measured time delays of 420.6 $\pm$ 1.8 days and 419.2 $\pm$ 2.2 days, respectively. Taking the weighted average of the two results, we find the time delay to be 420.0 $\pm$ 1.4 days. The reported delay listed is the weighted average of the two delays found by \citet{Shalyapin2012}.
\\
\item[--] {\it RX J1131$-$1231:} \citet{Tewes2013b} measured time delays between all pairs of light curves using three different numerical techniques. The time delay value listed in the table for each pair of light curves is for the technique that resulted in the smallest uncertainty. 
\\
\item[--] {\it H1413+117:} The light curves presented in \citet{Goicoechea2010} span less than one season and display poor variability. Hence our time delay measurements for the pairs AB, AC, and AD although in good agreement with the reported values, are of low precision and we could not reliably measure time delays for the pairs BC and BD. 
\\
\item[--] {\it CLASS B1600+434:} From both the optical light curves presented in \citet{Burud2000} (and made publicly available by \citet{Paraficz2006}) and the radio light curves presented in \citet{Koopmans2000}, we were unable to make a reliable time delay measurement. Although the optical light curves show good variability, they suffer from poor sampling and thus exclude the possibility of convincingly matching the variability features between light curves A and B. The radio light curves spanning one season is well sampled; however,   light curve A displays short time scale fluctuations that are not seen in   light curve B, thus making it difficult to measure the time delay unambiguously.
\\
\item[--] {\it HE 2149$-$2745:} We used the light curves presented in \citet{Burud2002a}, which were made publicly available by \citet{Paraficz2006}.
\end{description}

\begin{table*}
\caption{Our measurements of the time delays of HE 0435$-$1223 from two independent datasets.}
\label{table:HE0435}
\begin{center}
\begin{tabular}{l c c c}
\hline
\hline
Time delay & \citet{Courbin2011}\tablefootmark{a} & \citet{Blackburne2014} & Weighted average \\
           & (days)                               & (days)                 & (days) \\
\hline
$\Delta t_{AB}$ & 8.4 $\pm$ 1.4 & 12.3 $\pm$ 1.9 & 9.8 $\pm$ 1.1 \\
\hline 
$\Delta t_{AC}$ & 3.6 $\pm$ 3.4 & 2.7 $\pm$ 2.9 & 3.1 $\pm$ 2.2 \\
\hline 
$\Delta t_{AD}$ & 13.1 $\pm$ 1.1 & 15.8 $\pm$ 2.1 & 13.7 $\pm$ 1.0 \\
\hline 
$\Delta t_{BC}$ & $-$8.3 $\pm$ 1.5 & $-$7.7 $\pm$ 1.4 & $-$8.0 $\pm$ 1.0 \\
\hline 
$\Delta t_{BD}$ & 5.7 $\pm$ 1.7 & 7.9 $\pm$ 3.2 & 6.2 $\pm$ 1.5 \\
\hline 
$\Delta t_{CD}$ & 13.0 $\pm$ 1.1 & 14.1 $\pm$ 1.1 & 13.6 $\pm$ 0.8 \\
\hline 
\end{tabular}
\end{center}
\tablefoot{
\tablefoottext{a}{The SMARTS data points were excluded from the light curves of \citet{Courbin2011} so that the measured time delay values were independent of those measured from the SMARTS monitoring light curves of \citet{Blackburne2014} (see discussion in Sect. \ref{section:application}).}
}
\end{table*}

\section{$H_0$ from pixellated modeling of ten gravitational lenses}
\label{section:lens-modelling}

Of the 24 systems analyzed in the last section, 14 of them had light curves of sufficiently good quality to enable the measurement of at least one time delay between the images, adjacent to each other in terms of arrival-time order, to a precision of better than 20\% (which corresponds to a 5$\sigma$ detection of time delay). The ten systems which did not satisfy this criterion are Q0142$-$100 (UM673), FBQ 0951+2635, PG 1115+080, H1413+117, JVAS B1422+231, CLASS B1600+434, SDSS J1650+4251, PKS 1830$-$211, HE 2149$-$2745, and HS 2209+1914.

Of the 14 remaining systems, we did not model the mass distribution for four of them for the following reasons. SDSS J1001+5027 and SDSS J1206+4332 do not have accurate astrometric data measured from Hubble Space Telescope (HST) images or ground-based imaging with adaptive optics. Although the astrometry of JVAS B0218+357, which has a small image separation of 0.33$\arcsec$, has been measured from HST images by \citet{Sluse2012}, the authors warn about possibly large systematic errors in the published astrometry. SDSS J1029+2623 is a three-image cluster lens with highly complex mass distribution \citep[see][]{Oguri2013} and hence not amenable to lens-modeling following the simplistic approach described below. 

To perform mass-modeling of the remaining ten systems -- HE 0435$-$1223, SBS 0909+532, RX J0911.4+0551, Q0957+561, SDSS J1004+4112, HE 1104$-$1805, RX J1131$-$1231, SBS 1520+530, CLASS B1608+656 and WFI J2033$-$4723 -- to infer $H_0$, we used the publicly available PixeLens\footnote{\url{http://www.physik.uzh.ch/~psaha/lens/pixelens.php}} code \citep{Saha2004}, which builds an ensemble of pixellated mass maps compatible with the input data for a given system, which is comprised of the redshifts of the quasar and the lensing galaxy, the arrival-time order of the images, their astrometry relative to the center of the main lensing galaxy, and the known time delays between the images adjacent to each other in terms of arrival-time order. In case of quadruple lenses in which only some of the time delays are known, it is still possible to guess the arrival-time order of the images by following certain simple rules \citep[see][]{Saha2003}. 

We model all lenses, except SDSS J1004+4112, such that their mass profiles have inversion symmetry about the lens center, including any companion galaxy to the main lensing galaxy as a point mass. The lensing cluster in SDSS J1004+4112 consists of several galaxies besides the main lensing galaxy \citep[see][]{Inada2005} and hence was modeled without assuming inversion symmetry about the lens center.  

PixeLens builds models such that their projected density profiles are steeper than $|\boldsymbol{\theta}|^{-\gamma_{min}}$, where $|\boldsymbol{\theta}|$ is the distance from the center of the lens in angular units, the default value of $\gamma_{min}$ being 0.5. This is based on the observation that the total density distribution in the central regions of elliptical galaxies is close to isothermal (i.e., $r^{-2}$) and also the observation that the total density in the center of our Galaxy scales is $r^{-1.75}$ \citep[see][Sect. 2.2 and references therein]{Saha2004}. The profiles $r^{-2}$ and $r^{-1.75}$ correspond to projected density profiles of $|\boldsymbol{\theta}|^{-1}$ and $|\boldsymbol{\theta}|^{-0.75}$, respectively, in the special case of spherical symmetry. In this work, we relax the restriction of $\gamma_{min}$ = 0.5 and set $\gamma_{min}$ = 0 for those lenses in our sample in which the largest angular separation between the images is greater than 3$\arcsec$. The lenses in our sample that satisfy this criterion are RX J0911.4+0551, Q0957+561, SDSS J1004+4112, HE 1104$-$1805, and RX J1131$-$1231. A large image separation implies that there is significant lensing action from the cluster of which the main lensing galaxy is part, in which case the projected density profile can be shallower than $|\boldsymbol{\theta}|^{-0.5}$. 

For each system, we build an ensemble of 100 models, corresponding to 100 values of $H_0$. The mean of the 100 values gives the best estimate of $H_0$, the uncertainty of which is the standard deviation of the 100 values. This uncertainty includes only the uncertainty in the mass model. PixeLens assumes that the uncertainty in the input priors to be negligibly small, which is a reasonable assumption for the redshifts, if they are spectroscopically measured, and astrometry, if measured from HST or ground-based adaptive optics imaging. However, the measured time delays have finite uncertainties, which need to be propagated into the uncertainty of the estimated $H_0$. We do this by remodeling  each system after perturbing the time delay by its 1$\sigma$ uncertainty and noticing the deviation of the resulting value of $H_0$ from the original value. For high-precision time delays, the deviation in $H_0$ was found to be the same whether the delays were perturbed upward or downward. In general, the deviation in $H_0$ was found to be slightly larger when the delays were perturbed downward than when they were perturbed upward. Hence in this work, to get a conservative estimate of the contribution of the time delay uncertainty to the uncertainty in $H_0$, we decrease the time delay by its 1$\sigma$ uncertainty and find the resulting increase in $H_0$. This uncertainty in $H_0$ resulting from the time delay uncertainty is added in quadrature to the uncertainty in $H_0$ resulting from mass modeling to find the total uncertainty. In the case of quadruple lenses where more than one time delay is known, we perturb each delay individually while leaving the other delays unchanged to infer its uncertainty contribution. The uncertainty contribution from each independent time delay is then added in quadrature to the uncertainty in $H_0$ resulting from the uncertainty in the mass model to find the total uncertainty.  

In order to include the effects of external shear, an approximate direction of the shear axis needs to be specified and PixeLens will search for solutions within 45$\degr$ of the specified direction. Since there is no simple rule to guess the direction of the external shear for a given system, for each system, we repeated the modeling specifying the approximate direction of the shear axis as 90$\degr$, 45$\degr$, 0$\degr$, and $-$45$\degr$ (in this instance, specifying $\theta$ and $\theta$+180$\degr$ are equivalent). We thus obtain four estimates of $H_0$ and their uncertainties. In each case, we propagate the uncertainty contributions from the known time delays to the uncertainty in $H_0$, as discussed previously. The final estimate of $H_0$ and its uncertainty are found using maximum likelihood analysis, optimizing their values so as to maximize the joint posterior probability of these two parameters for the sample consisting of the four $H_0$ values and their uncertainties \citep[see][Eq. 7]{Barnabe2011}. In optimizing the value of the uncertainty, we choose the minimum limit to be the smallest of the four uncertainties. We note here that for the system HE 1104$-$1805, the choices of the approximate direction of the shear axis of 90$\degr$ and $-$45$\degr$ were found to lead to unphysical models involving negative values in the mass pixels. Hence for this system, the maximum likelihood analysis was carried out using only the two $H_0$ values resulting for the approximate shear directions of 45$\degr$ and 0$\degr$.  

\begin{figure}
\begin{center}
\resizebox{1.0 \hsize}{!}{\includegraphics{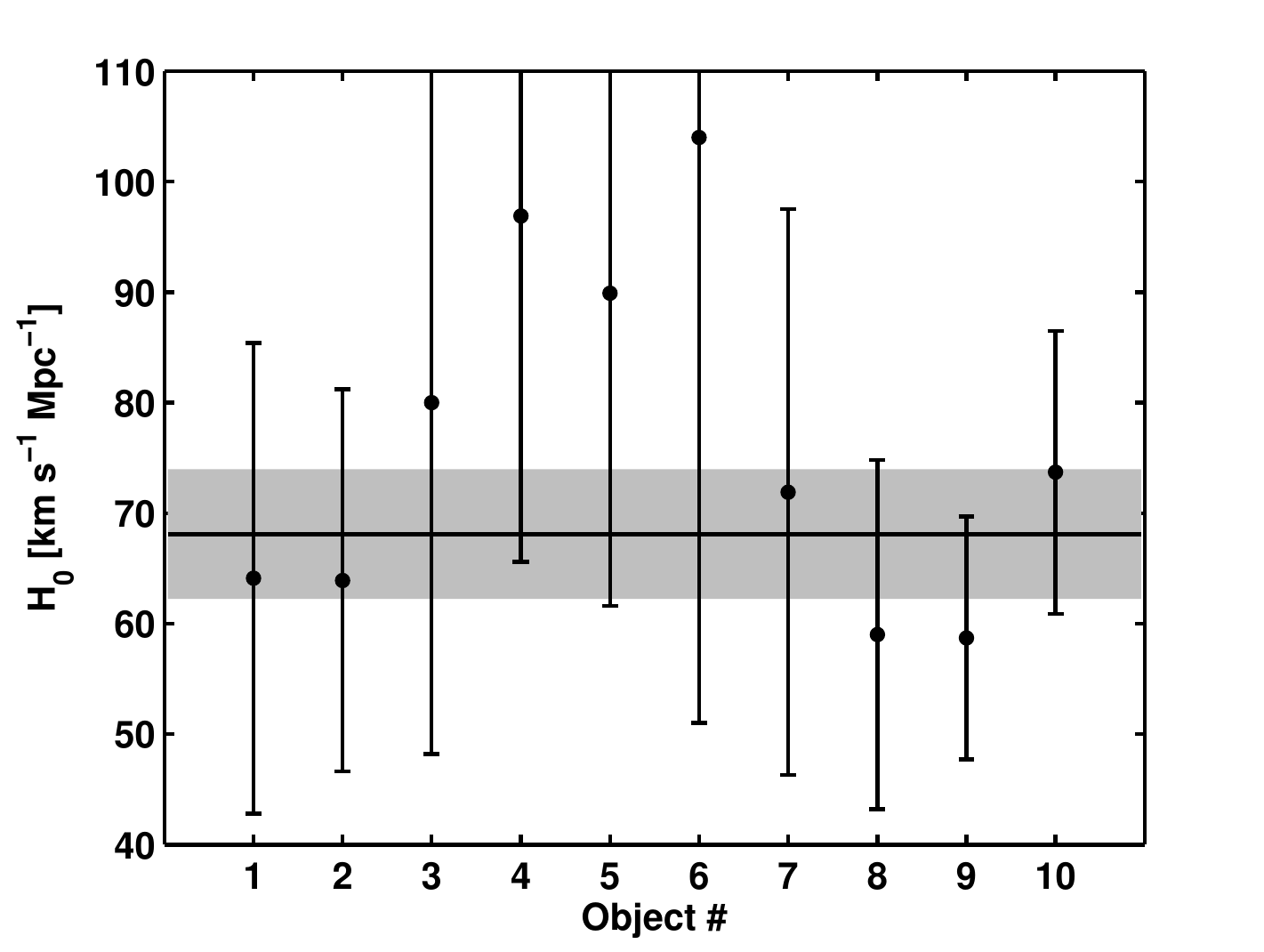}}
\caption{The $H_0$ estimates and their 1$\sigma$ uncertainties for the ten gravitational lenses -- {\it (1)} HE 0435$-$1223, {\it (2)} SBS 0909+532, {\it (3)} RX J0911.4+0551, {\it (4)} Q0957+561, {\it (5)} SDSS J1004+4112, {\it (6)} HE 1104$-$1805, {\it (7)} RX J1131$-$1231, {\it (8)} SBS 1520+530, {\it (9)} CLASS B1608+656, and {\it (10)} WFI J2033$-$4723. The best estimate of $H_0$ and its 1$\sigma$ confidence interval, inferred through maximum-likelihood analysis, are represented by the horizontal line and the gray shaded region, respectively.}
\label{figure:hubble-constant}
\end{center}
\end{figure}

The input priors for each system and the resulting $H_0$ estimates are summarized in Table \ref{table:lens-modelling}. In Fig. \ref{figure:hubble-constant} we plot the $H_0$ estimates from the ten lenses, all of which are seen to agree with each other within their error bars. To combine the ten independent estimates into a best estimate of $H_0$, we again employ maximum likelihood analysis, as described above. However, in this case, in optimizing the value of the uncertainty of the best estimate of $H_0$, the minimum limit is chosen to be the uncertainty of the weighted average of the ten values. We infer a value of $H_0$ of 68.1 $\pm$ 5.9 km s$^{-1}$ Mpc$^{-1}$ (1$\sigma$ uncertainty, 8.7\% precision) for a spatially flat universe having $\Omega_m$ = 0.3 and $\Omega_\Lambda$ = 0.7. The reason for employing maximum likelihood analysis in this case, rather than taking a simple weighted average is to detect the presence of any unmodeled uncertainties. However, as can be seen from Fig. \ref{figure:hubble-constant}, the $H_0$ estimates from the individual systems all agree with each other within their error bars and hence the $H_0$ value inferred above through maximum likelihood analysis is only marginally different from the weighted average. For the source and lens redshifts of the current sample, we find the $H_0$ estimate to decrease by 7.1\% for the Einstein-de Sitter universe ($\Omega_m$ = 1.0 and $\Omega_\Lambda$ = 0.0) and increase by 2.3\% for an open universe having $\Omega_m$ = 0.3 and $\Omega_\Lambda$ = 0.0, thus illustrating the low level of dependence of the inferred value of $H_0$ on the precise values of $\Omega_m$ and $\Omega_\Lambda$. In Table \ref{table:lens-modelling}, we also list the $H_0$ estimates obtained without propagating the time delay uncertainties. We see that the dominant contribution to uncertainty in $H_0$ results from the uncertainty in the mass model.   

\begin{table*}
\caption{Summary of input data to PixeLens and resulting $H_0$ estimates.}
\label{table:lens-modelling}
\begin{center}
\begin{tabular}{l l c r r l c l}
\hline
\hline
Object & Redshifts & Image\tablefootmark{a} & $\Delta$RA\tablefootmark{c} & $\Delta$Dec\tablefootmark{c} & Delay\tablefootmark{d} & $H_0$\tablefootmark{e} & References\tablefootmark{f} \\
       &           & / P.M.\tablefootmark{b}& ($\arcsec$)                 & ($\arcsec$)                  & (days) & (km s$^{-1}$ Mpc$^{-1}$) &        \\
\hline
HE 0435$-$1223 & $z_l$ = 0.4546 & A & 1.1706 & 0.5665 &              & 64.1 $\pm$ 21.3 & \citet{Morgan2005} \\
                     & $z_s$ = 1.689  & C & $-$1.2958 & $-$0.0357 &        & (64.1 $\pm$ 19.4) & \citet{Wisotzki2002} \\
                     &                      & B & $-$0.3037 & 1.1183 & 8.0 $\pm$ 1.0 &             & \citet{Courbin2011} \\
                     &                      & D & 0.2328 & $-$1.0495 &                     &             &                           \\
\hline
SBS 0909+532 & $z_l$ = 0.830 & B & 0.5228    & $-$0.4423 &                      & 63.9 $\pm$ 17.3 & \citet{Lubin2000} \\
             & $z_s$ = 1.377 & A & $-$0.4640 & 0.0550    & 45.9 $\pm$ 3.1 & (63.9 $\pm$ 16.8)     & \citet{Kochanek1997} \\
             &               &   &           &           &                      &                       & \citet{Sluse2012} \\
\hline
RX J0911.4+0551 & $z_l$ = 0.769 & B             & $-$2.2662 & 0.2904    &                        & 80.0 $\pm$ 31.8   & \citet{Kneib2000} \\
                & $z_s$ = 2.800 & A2            & 0.9630    & $-$0.0951 & 141.9 $\pm$ 12.3 & (80.0 $\pm$ 31.0) & \citet{Bade1997} \\
                &               & A1            & 0.7019    & $-$0.5020 &                        &                         & \citet{Sluse2012} \\
                &               & A3            & 0.6861    & 0.4555    &                        &                         &                   \\
                &               & P.M.          & $-$0.7582 & 0.6658    &                        &                         &                   \\
\hline
Q0957+561 & $z_l$ = 0.361 & A                & 1.408 & 5.034    &                                & 96.9 $\pm$ 31.3   & \citet{Walsh1979} \\
          & $z_s$ = 1.41  & B                & 0.182 & $-$1.018 & 420.0 $\pm$ 1.4          & (96.9 $\pm$ 31.3) & \citet{Fadely2010} \\
\hline
SDSS J1004+4112 & $z_l$ = 0.68  & C                & 3.925    & $-$8.901 &                       & 89.9 $\pm$ 28.3   & \citet{Oguri2004} \\
                & $z_s$ = 1.734 & B                & $-$8.431 & $-$0.877 & 777.1 $\pm$ 9.2 & (89.9 $\pm$ 28.1) & \citet{Inada2003} \\
                &               & A                & $-$7.114 & $-$4.409 & 37.2 $\pm$ 3.1  &                         & \citet{Inada2005} \\
                &               & D                & 1.285    & 5.298    &                        &                         &                   \\
\hline
HE 1104$-$1805 & $z_l$ = 0.729 & B              & 1.9289 & $-$0.8242 &                               & 104.0 $\pm$ 53.0   & \citet{Lidman2000} \\
               & $z_s$ = 2.319 & A              & $-$0.9731 & 0.5120 & 157.1 $\pm$ 3.6         & (104.0 $\pm$ 52.9) & \citet{Smette1995} \\
               &               &                &           &        &                               &                          & \citet{Sluse2012} \\
\hline
RX J1131$-$1231 & $z_l$ = 0.295 & C             & $-$1.460 & $-$1.632 &                              & 71.9 $\pm$ 25.6   & \citet{Sluse2003} \\
                & $z_s$ = 0.658 & B             & $-$2.076 & 0.662    &                              & (71.9 $\pm$ 25.6) & \citet{Suyu2013} \\
                &               & A             & $-$2.037 & $-$0.520 &                              &                         &                  \\
                &               & D             & 1.074    & 0.356    & 91.7 $\pm$ 0.7         &                         &                  \\
                &               & P.M.          & $-$0.097 & 0.614    &                              &                         &                  \\
\hline
SBS 1520+530 & $z_l$ = 0.761 & A                & $-$1.1395 & 0.3834    &                            & 59.0 $\pm$ 15.8   & \citet{Auger2008} \\
             & $z_s$ = 1.855 & B                & 0.2879    & $-$0.2691 & 124.2 $\pm$ 8.1      & (59.0 $\pm$ 15.3) & \citet{Chavushyan1997} \\
             &               &                  &           &           &                            &                         & \citet{Sluse2012} \\
\hline
CLASS B1608+656 & $z_l$ = 0.6304 & B            & 1.2025    & $-$0.8931 &                            & 58.7 $\pm$ 11.0   & \citet{Myers1995} \\
                & $z_s$ = 1.394  & A            & 0.4561    & 1.0647    & 32.4 $\pm$ 3.0       & (58.7 $\pm$ 10.8) & \citet{Fassnacht1996} \\
                &                & C            & 1.2044    & 0.6182    &                            &                         & \citet{Sluse2012} \\
                &                & D            & $-$0.6620 & $-$0.1880 & 41.3 $\pm$ 1.6       &                         &                   \\
                &                & P.M.         & 0.7382    & 0.1288    &                            &                         &                   \\
\hline
WFI J2033$-$4723 & $z_l$ = 0.661 & B            & 1.4388    & $-$0.3113 &                            & 73.7 $\pm$ 12.8   & \citet{Eigenbrod2006} \\
                 & $z_s$ = 1.66  & A1           & $-$0.7558 & 0.9488    & 37.6 $\pm$ 2.1       & (73.3 $\pm$ 11.6) & \citet{Morgan2004} \\
                 &               & A2           & $-$0.0421 & 1.0643    &                            &                         & \citet{Vuissoz2008} \\
                 &               & C            & $-$0.6740 & $-$0.5891 & 23.6 $\pm$ 2.5       &                         &                     \\
\hline
Combined        &               &                &           &           &                           & 68.1 $\pm$ 5.9   &              \\
                &               &                &           &           &                           & (67.9 $\pm$ 5.6) &              \\
\hline
\end{tabular}
\end{center}
\tablefoot{
\tablefoottext{a}{The QSO images are listed in arrival-time order.}
\tablefoottext{b}{`P.M.' is the abbreviation for point mass and refers to secondary lensing galaxies.}
\tablefoottext{c}{The astrometry of the QSO images and point masses are specified with respect to the center of the main lensing galaxy.}
\tablefoottext{d}{The time delay of a given image is listed (if measured to a precision  better than 20\%) with respect to the previous image in terms of arrival-time order.}
\tablefoottext{e}{In parentheses we provide the $H_0$ estimates and their uncertainties without propagating the uncertainties in time delays.}
\tablefoottext{f}{The references are listed for measurements of lens redshift ($z_l$), source redshift ($z_s$), and astrometry.}
}
\end{table*} 

\section{Conclusion}
\label{section:conclusion}
We have presented a homogeneous curve-shifting analysis of the light curves of 
24 gravitationally lensed quasars for which time delays have been reported in 
the literature  so far. Time delays were measured using the 
difference-smoothing technique and their uncertainties were estimated using 
realistic simulated light curves; a recipe for creating these light curves  with  known time delays in a plausible range around the measured
delay was introduced 
in this work. We identified 14 systems to have light curves of sufficiently good 
quality to enable the measurement of at least one time delay between the 
images, adjacent to each other in terms of arrival-time order, to a precision 
of better than 20\% (including systematic errors). Of these 14 systems, we 
performed pixellated mass modeling using the publicly available PixeLens 
software for ten of them, which have known lens redshifts, accurate astrometric 
information, and sufficiently simple mass distributions, to infer the value 
of $H_0$ to be 68.1 $\pm$ 5.9 km s$^{-1}$ Mpc$^{-1}$ (1$\sigma$ 
uncertainty, 8.7\% precision) for a spatially flat universe 
having $\Omega_m$ = 0.3 and $\Omega_\Lambda$ = 0.7. We note here that we have followed a relatively simple lens modeling approach to constrain $H_0$ and our analysis does not account for biases resulting from line-of-sight effects.  

Our measurement closely matches  a recent estimate of $H_0$ = 69.0 $\pm$ 6 (stat.) $\pm$ 4 (syst.) 
km s$^{-1}$ Mpc$^{-1}$ found by \cite{Sereno2014} using a method 
based on free-form modeling 
of 18 gravitational lens systems.  Our value is 
also consistent with the recent measurements of 
$H_0$ by \citet{Riess2011}, \citet{Freedman2012} and \citet{Suyu2013}; however, it has lower
precision. Increasing the number of lenses with good-quality light curves, accurate astrometry, and known lens redshift
from the current ten used in this study can bring down the uncertainty in 
$H_0$. 

In the future such high-precision time delays will become
available from projects such as COSMOGRAIL \citep{Tewes2012} involving dedicated medium-sized telescopes. 
In addition, the next generation of cosmic surveys such as the Dark Energy Survey
(DES), the Large Synoptic Survey Telescope (LSST; \citealt{Ivezic2008}), and the Euclid mission
will detect a large sample of lenses, and time delays might be available for 
a large fraction of them,  consequently enabling measurement of $H_0$
to an accuracy better than 2\%. Furthermore, detection of gravitational wave
signals from short gamma-ray bursts associated with neutron star binary mergers
in the coming decade could constrain $H_0$ to better than 1\% \citep{Nissanke2013}.

\begin{acknowledgements}
We thank Jim Lovell for providing us with the light curves of PKS 1830$-$211. We acknowledge useful discussions with Malte Tewes, G. Indu, Leon Koopmans, 
Prashanth Mohan and Matthias Bartelmann. We thank Frederic Courbin and Georges Meylan for carefully reading the manuscript and offering helpful comments.
We thank the organizers of the Strong Lens Time Delay Challenge for enabling a blind test of our algorithm and subsequently providing  the truth files, which helped us to refine our curve-shifting procedure. We thank the anonymous referee for constructive reports  that helped to improve the presentation of this work. 
\end{acknowledgements}

\bibliographystyle{aa}
\bibliography{h0_biblio}
\end{document}